# Parameters for a 30 GeV Undulator Test Facility in the FFTB/LCLS *

P. Krejcik
Stanford Linear Accelerator Center, Stanford University, Stanford CA 94309

* Work supported by Department of Energy contract DE–AC03–76SF00515.

# Parameters for a 30 GeV Undulator Test Facility in the FFTB/LCLS

Patrick Krejcik*

## Abstract

The parameters for a 30 GeV test beam are outlined for use with an undulator in the FFTB tunnel where the LCLS will eventually be housed. It is proposed to use the SLAC linac and damping rings in their present mode of operation for PEP II injection, where 30 GeV beams are also delivered at 10 Hz to the FFTB. High peak currents are obtained with the addition of a second bunch compressor in the linac. In order to minimize the synchrotron radiation induced emittance growth in the bunch compressor it is necessary to locate the new bunch compressor at the low-energy end of the linac, just after the damping rings. The bunch compressor is a duplicate of the LCLS chicane-style bunch compressor. This test beam would provide an exciting possibility to test LCLS undulator sections and provide a unique high-brightness source of incoherent X-rays and begin developing the LCLS experimental station. The facility will also act as a much needed accelerator test bed for the production, diagnostics and tuning of very short bunches in preparation for the LCLS after the photo injector is commissioned.

* email pkr@slac.stanford.edu

# 1 Introduction

There are two factors that motivate building this test facility with the existing 30 GeV beams in the SLAC linac. One is a desire from the user community to develop high-brightness sources of X-rays as soon as possible. The LCLS promises to deliver coherent X-rays in the 1.5 to 15 Angstrom range with a brightness that is 10 orders of magnitude higher than existing undulator insertion devices. An undulator using the 30 GeV beam at the end of the linac would provide only incoherent X-rays in this range, but their brightness would still be several orders of magnitude above existing sources. This makes the proposed facility a stepping stone towards developing experiments for the LCLS. Not only would the experiments act as prototypes for experiments with the much brighter LCLS beams, but they would also be built up within the future LCLS user facility area.

The prototyping and testing of LCLS accelerator technology is the second principal motivating factor for the early development of this intermediate-brightness X-ray source. In parallel with the present development of the photo injector for the LCLS it is possible to use the existing SLAC linac beam from the damping ring to produce, diagnose and tune very short bunches. The preservation of the emittance during bunch manipulation and acceleration in the SLAC linac is a key milestone to be demonstrated for the successful operation of the LCLS.

The beam from the SLAC damping rings falls short of the final LCLS requirements, but the small vertical emittance from the damping rings, which is comparable to the LCLS transverse emittance requirements, is an ideal diagnostic for investigating accelerator issues. The present Ring-To-Linac, RTL, compressor is not sufficient to produce the kilo-ampere peak currents desirable in an X-ray undulator, so a second bunch compressor must be installed in the linac.

Using the SLAC linac for LCLS accelerator development has the following advantages:

- The bunch compressor has identical beam line and rf parameters as the LCLS design and is installed in the same linac housing as the LCLS.
- Bunch compression and emittance growth measurements would be performed with the same accelerating structures and hence wakefields as the LCLS
- The complete operation of the linac bunch compressor system would be tested, including the controls and instrumentation for reproducible tuning.

- Phase stability and pulse-to-pulse jitter of the LCLS linac would also be tested in this facility. An advance start could be made on the precision phase measurement and stabilization R&D needed for the LCLS.

Finally, the promotion of research into the production and tuning of very short bunches fits in well with our laboratory's broader program and goals. The shorter bunches delivered to the end of the linac could be immediately used to advantage in the advanced accelerator research experiments into plasma wakefield acceleration (E150). The production, diagnostics and tuning of short bunches is directly applicable to the Next Linear Collider development where the same issues of tuning and stability are also worthy of study.

# 2 Accelerator Layout

The basis of this proposal is to use the existing SLAC linac in a configuration as close as possible to the present setup for PEP II injection, as shown in figure 1. The North Damping Ring provides 10 Hz electrons for PEP II injection and positron production on separate 1/60$^{th}$ of a second duration store cycles. The PEP II injection bunch is extracted from the linac at the 9.5 GeV point in sector 10 where it is deflected into the NIT bypass line by a 30 Hz pulsed magnet. The scavenger bunch for positron production is accelerated up to sector 19 to approximately 30 GeV. In sector 4 a pulsed magnet deflects positrons at 3.5 GeV into the SIT bypass line. The damping rings operate at 30 Hz so a 3$^{rd}$ FFTB beam pulse is also available for acceleration to the end of the linac. Beyond sector 19 the beam coasts through non-energized sectors to sector 30. Sector 30 is powered for energy feedback control of the beam. It is proposed to use this beam pulse with an undulator installed in the FFTB tunnel to produce X-rays for synchrotron light users.

The damping ring complex uses a bunch compressor in the ring-to-linac (RTL) beam line to compress the 6 mm bunch from the damping ring to approximately 1.3 mm at the entrance to sector 2 of the linac. A second bunch compression stage is therefore required to reach the higher peak currents that are of interest to X-ray production.

In an earlier note [Emma, Krejcik] we described the parameters for a chicane style bunch compressor to be installed in sector 24. The basis of this earlier proposal was to build one of the LCLS bunch compressors at the location prescribed for future operation with the LCLS RF photo-injector. A 4.5 GeV test beam was to be delivered from the damping rings to sector 24 (by acceleration to 9.5 GeV and subsequent deceleration to 4.5 GeV) in order to test the bunch compressor using the small vertical emittance from the damping ring as a probe. This scenario is not suitable for use with the higher energy beams proposed here for X-ray production at 30 GeV. In order to minimize emittance growth due to synchrotron radiation the bunch compression must be done at low energy. Alternatively, at high energy a very large radius of curvature bend must be employed. An example would be the SLC arcs, but this is a different technology and takes the beam out of the LCLS experimental area. A separate proposal by Emma and Frisch (SLAC Pub 8308) describes such a scheme.

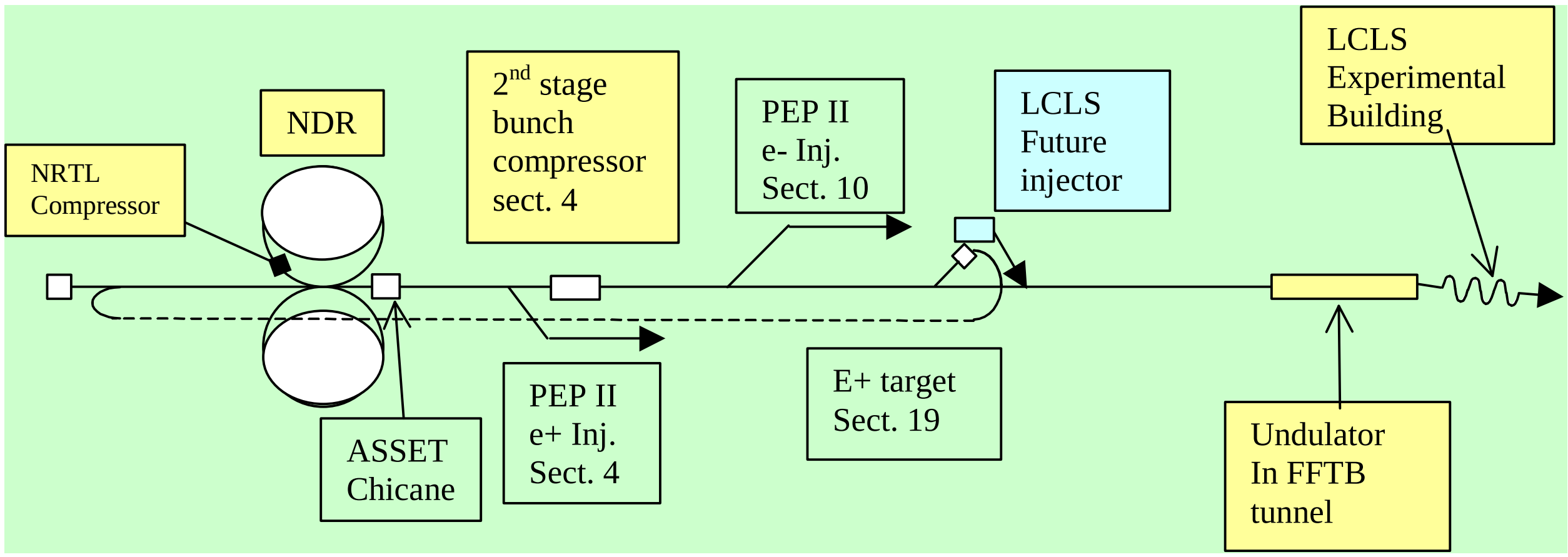

**Figure 1** Schematic of the SLAC linac showing the second stage bunch compression in sector 4 proposed for producing short bunches in an undulator located in the FFTB tunnel

In order to make use of the beam with an undulator in the LCLS experimental area beyond the FFTB tunnel it will be necessary to compress the beam at the low energy end of the linac. Several locations can be considered that are suitable for a chicane style bunch compressor. One possibility is in sector 2 where the 1.19 GeV damping ring beam is injected into the linac. The ASSET chicane and an instrumentation section are also located in sector 2. The new bunch compressor chicane has to be downstream of at least one girder of accelerating sections (i.e. one klystron) that would be appropriately phased to introduce a suitable correlated energy spread.

Another possibility is in sector 4 at the positron extraction chicane. At this location the beam is close in energy to the design energy of 4.5 GeV for the LCLS 2$^{nd}$ bunch compressor, so that an installation here could be considered a working prototype for the LCLS. The specific correlated energy spread required at the entrance to the bunch compressor chicane can easily be achieved with the phase of the klystrons in sectors 2 through 4. It might be tempting to consider modifying the existing positron extraction line with the addition of dc chicane bends to introduce the $R_{56}$ term into the electron transport line. However, there may be some conflict of interest in the choice of bend plane for the compressor and the positron extraction.

For this exercise it will be assumed that the new bunch compressor chicane will be located at 4.5 GeV in sector 4 beyond the positron extraction bend. The parameters for the 2nd LCLS bunch compressor system, designed by Paul Emma, can therefore be adopted for this study.

With the addition of the chicane in sector 4 some thought can be given to whether the normal operation of the PEP II injection bunch and the scavenger bunch would be interfered with. In the unlikely event that the shorter bunches are disruptive to PEP II injection or positron production then pulsed phase shifters can be used in sectors 2 through 4 to remove the correlated energy spread in only those bunches at sector 4. Without the correlated energy spread the chicane has no compressive effect.

# 3 Electron Beam Parameters

In table 1 the parameters at each of the LCLS bunch compressors are compared with the sector 4 parameters.

In addition to compressing the beam it is desirable to preserve as much of the low damping ring transverse emittance through to the end of the linac. The relatively long bunch length of the 4.5 GeV beam at the entrance to sector 4 compared to BC2 in the LCLS means that a larger correlated energy spread of 2.3% is required. Such a large energy spread is potentially damaging to emittance growth, but since the beam is further accelerated to 30 GeV the relative energy spread becomes less and the effect on emittance growth diminishes. This can be further verified with particle tracking.

Further compression of the damping ring beam beyond the second stage in sector 4 has little benefit because of the larger longitudinal emittance relative to the parameters of the RF photo injector proposed for LCLS.

| Parameter | LCLS BC1 | LCLS BC2 | SLC RTL | SLC sector 4 | SLC sector 30, | unit |
|---|---|---|---|---|---|---|
| Energy | 0.25 | 4.5 | 1.19 | 4.5 | 30 | GeV |
| Initial bunch length (rms) | 1 | 0.39 | 6 | 1.3 | - | mm |
| Final bunch length (rms) | 0.39 | 0.02 | 1.3 | 0.10 | 0.10 | mm |
| Final energy spread (rms) | 1.3 | 0.9 | 1.3 | 2.3 | 0.55 | % |
| Bunch charge | 1 | 1 | 1 | 1 | 1 | nC |
| Compressor $R_{56}$ | –31 | –27 | –605 | –56 | - | mm |
| Vertical emittance(norm.) | 1 | 1 | 1.5 | 2.0 | 2.0 | $\mu$m |
| Peak current | 0.77 | 15 | 0.23 | 1.2 | 1.2 | kA |

**Table 1**. Parameters of the LCLS compressors together with the RTL and sector 4 parameters.

## 3.1 Longitudinal tracking

Longitudinal tracking studies of the existing linac configuration with the RTL compressor are compared with the modified linac in which a chicane compressor is added in sector 4. The LITRACK program (MATLAB version written by Paul Emma, based on Karl Bane's program) is used to track 50,000 particles from the exit of the damping rings to the end of the linac. A gaussian distribution with a 6mm rms bunch length is assumed at the exit of the damping rings, although we know from experience that at higher currents the bunch does not remain gaussian in the damping rings and this effect can be included later. The longitudinal wakefield loading on the bunch is computed in the program from Karl Bane's point charge wake function for the SLAC linac structure.

In the case of the present linac configuration the RF amplitude of the compressor is set for the fully compressed state to give the shortest low-charge bunch at the end of the linac. The linac is phased for on-crest acceleration to 30 GeV at sector 20 and then coasts to sector 30, to simulate the present mode of FFTB operation. The longitudinal distribution and energy spread are show in figure 2 for the bunch at sector 30.

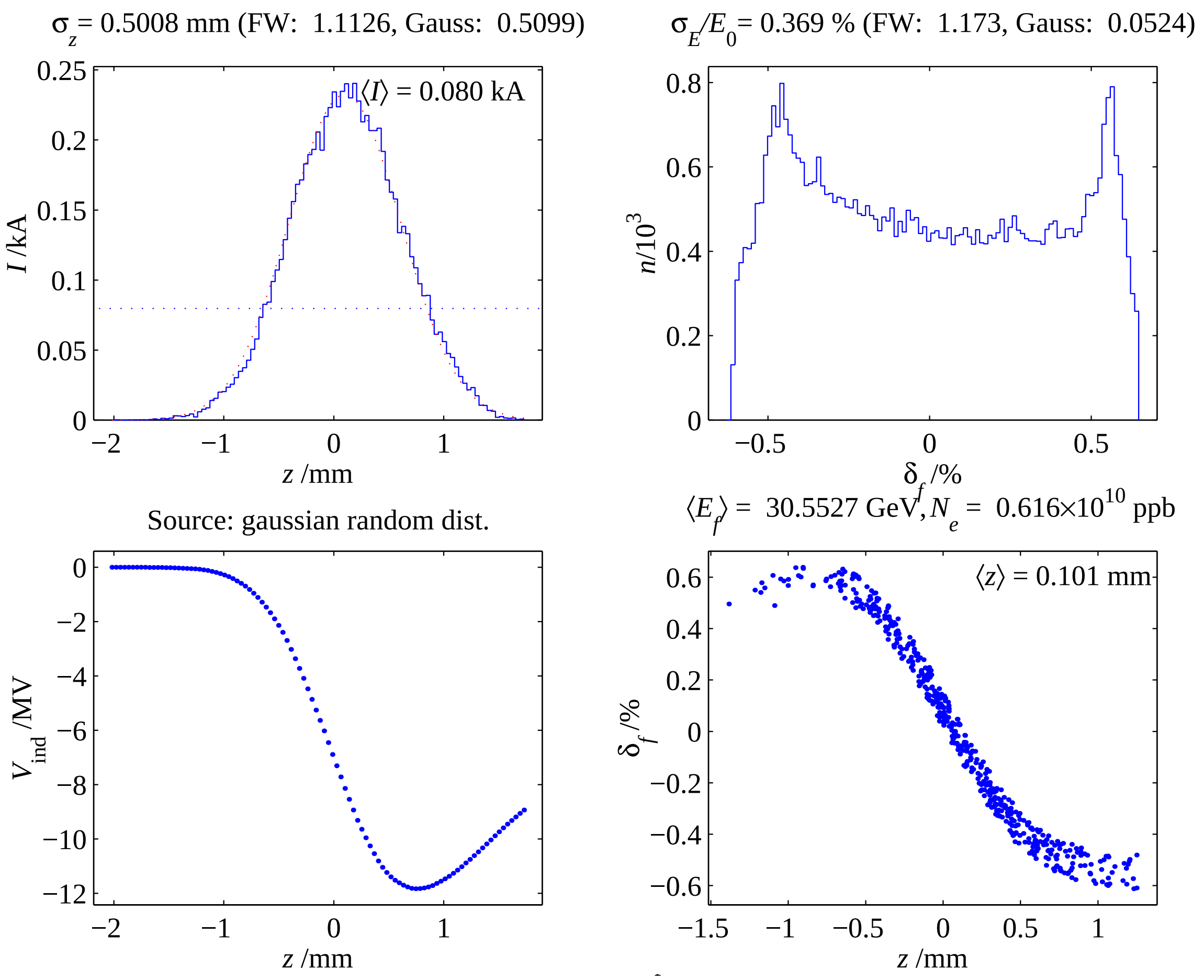

**Figure 2**: Sector 30 bunch length and energy distribution for a 1 nC bunch fully-compressed in the RTL, accelerated to 30 GeV in sector 20 and allowed to coast to the end of the linac

For operation with the additional bunch compressor in sector 4 the bunch is slightly over-compressed in the RTL (38 MV compressor RF amplitude compared to 34 MV for fully-compressed). The correlated energy spread of the over-compressed bunch helps compensate the wakefield-induced energy spread in the linac. Sectors 2 through 4 are phased off crest as indicated in table 2 to induce the correct phase-energy correlation for the given $R_{56}$ of the chicane in sector 4. Sectors 5 through 20 are phased to accelerate the beam on crest to 30 GeV. The remaining sectors 21 through 30 are idle so that only the wakefields effects the beam as it coasts to the end of the linac. The longitudinal distribution and energy spread are show in figure 3 for the bunch at sector 30.

As the charge in the bunch is increased the wakefields in the linac cause an increasing amount of correlated energy spread in the bunch by the time it reaches the end of the linac. This is illustrated in figure 4 where the bunch length, peak current and energy spread are plotted for the tracking results for 1 nC, 2 nC, 3 nC and 4 nC. In these simulations all linac parameters are held constant and only the charge is increased.

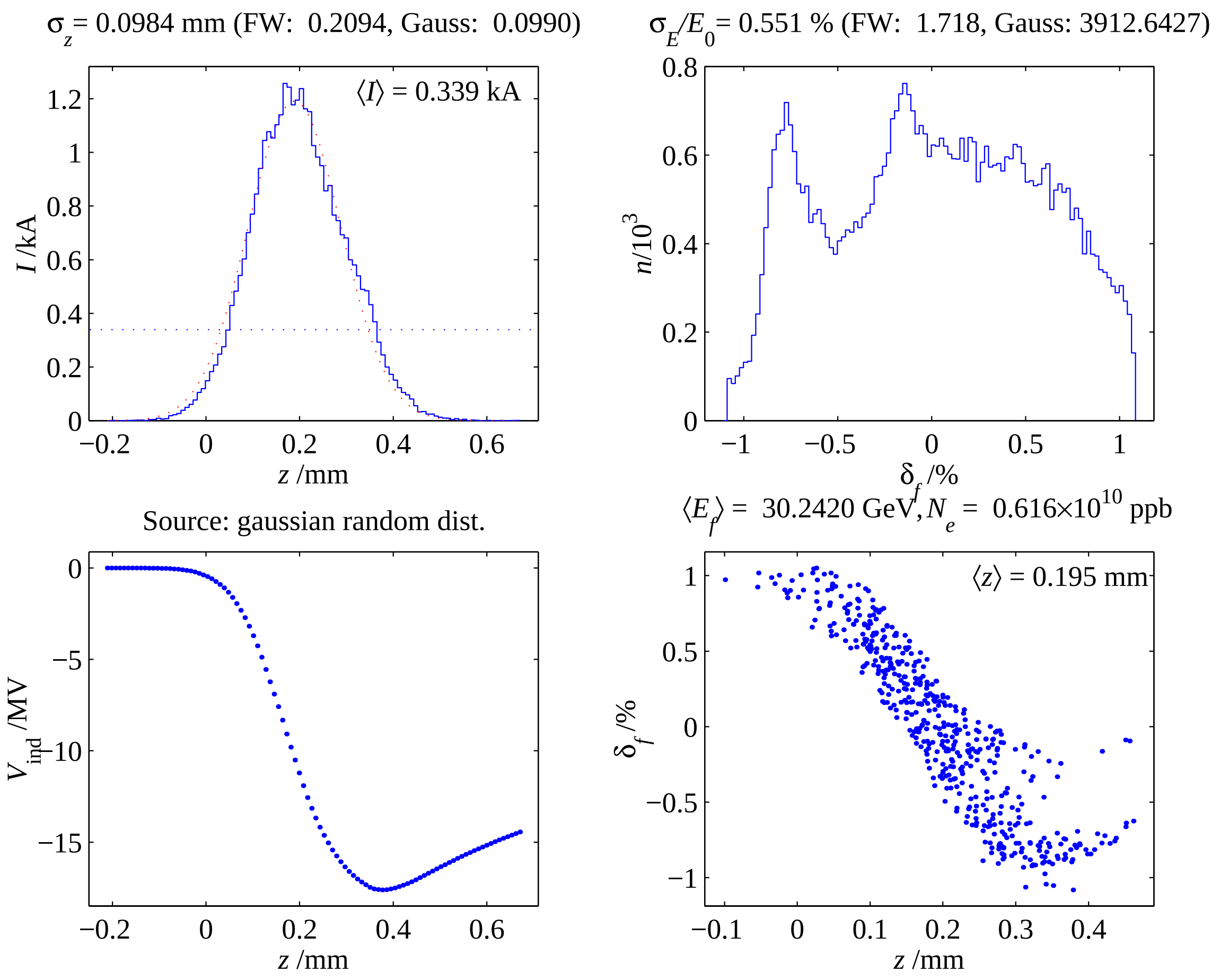

**Figure 3**: Sector 30 bunch length and energy distribution for a 1 nC bunch with second stage compression in sector 4, accelerated to 30 GeV in sector 20 and allowed to coast to the end of the linac.

| Parameter | symbol | value | unit |
|---|---|---|---|
| RTL RF Compressor amplitude stand alone | $V_{comp}$ | 34 | MV |
| RTL RF Compressor amplitude with sect 4 | $V_{comp}$ | 38 | MV |
| Average RF gradient used in sectors 2-4 | $\langle G_1\rangle$ | 12 | MV/m |
| Average RF gradient used in sectors 4-20 | $\langle G_2\rangle$ | 15.3 | MV/m |
| Average RF phase used in sectors 2-4 | $\langle \varphi_1\rangle$ | –22.0 | deg |
| Average RF phase used in sectors 4-20 | $\langle \varphi_2\rangle$ | –0 | deg |
| Energy in sector-4 | $E_1$ | 4.5 | GeV |
| Energy in sector-20 | $E_2$ | 30.0 | GeV |
| Relative energy spread in sector-4 (rms) | $\sigma_{\delta_1}$ | 2.3 | % |
| Relative energy spread in sector-30 (rms) | $\sigma_{\delta_2}$ | 0.55 | % |

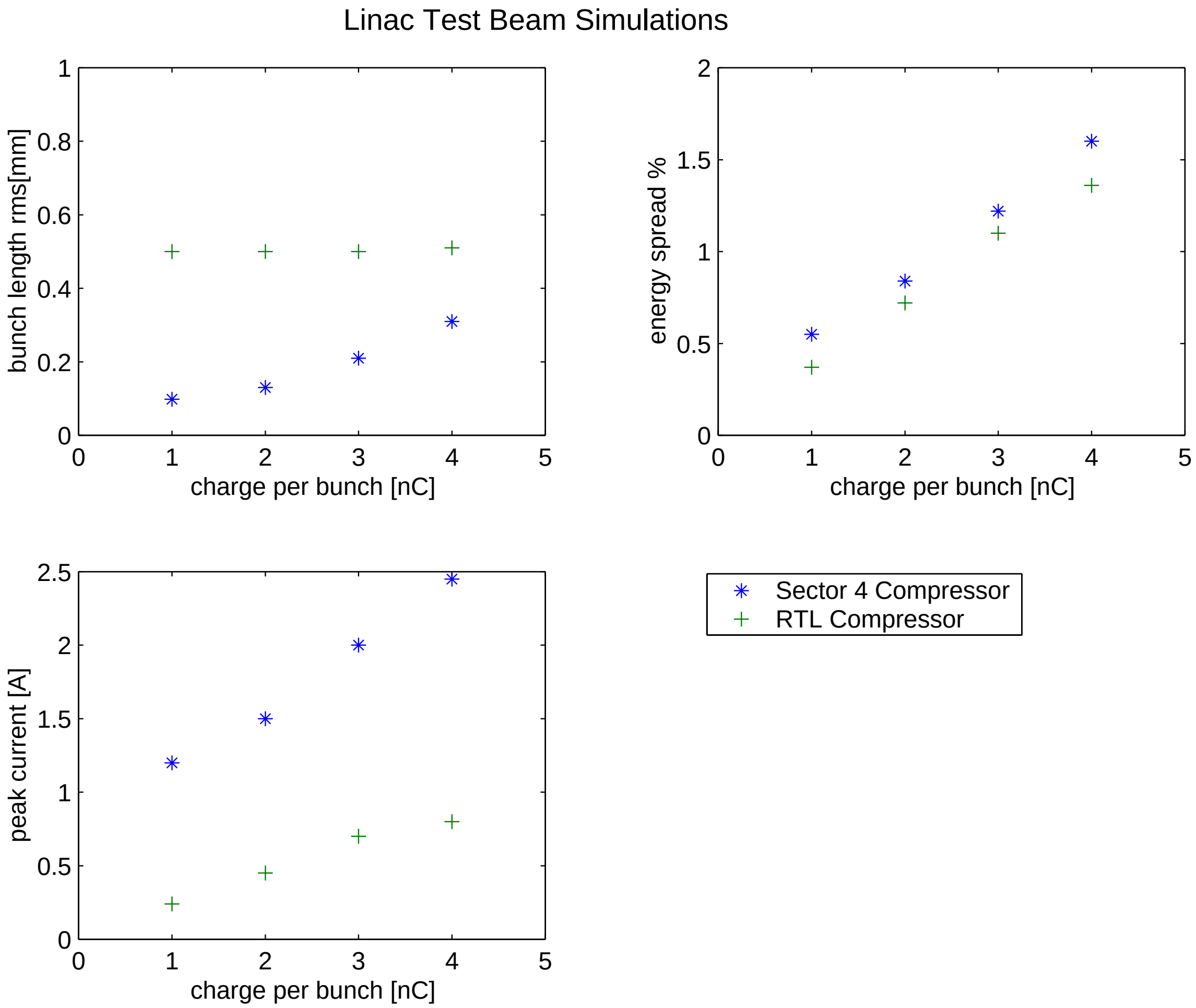

**Figure 4** Dependence on bunch charge for the bunch length, energy spread and peak current of the SLC beam (+) and the two-stage compressed beam (*).

# 4 Components

## 4.1 Magnet chicane

The design of the first chicane of the second bunch compressor in the LCLS calls for four 1.5 m long dipoles to provide the momentum compaction ($R_{56}$) term. The overall length of the chicane system is 13.2 m with a displacement from the accelerator axis of 0.28 m, as shown in figure 2. Four of the 10’ SLAC accelerating sections need to be removed to accommodate the chicane magnets. The LCLS design proposes that the inner two dipole be made movable so that the strength of the chicane can be varied to control the $R_{56}$.

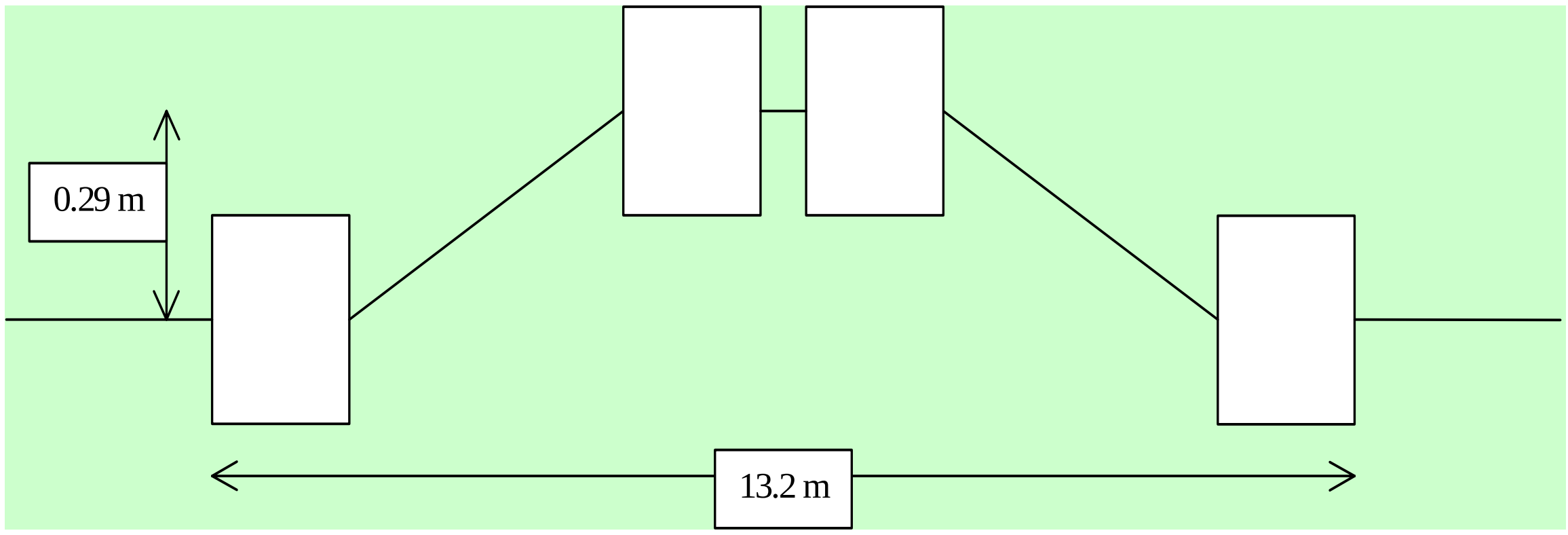

**Figure 5** A four-dipole chicane supplies the necessary momentum compaction for the second stage bunch compressor proposed in sector 4.

It would be preferable to orient the dipoles to deflect the beam vertically so that in-plane emittance growth phenomena could be studied more carefully using the small vertical emittance from the damping rings. This might be, in general, desirable for the LCLS to avoid blocking the tunnel access with horizontal bends. It is further foreseen to make the chicane deflection angle adjustable to control the value of $R_{56}$.

The cost of the bunch compressor section can be significantly reduced by re-using components from the SLC. Several magnets in the SLC have suitable parameters for use in the LCLS and have been listed in a separate note (P. Krejcik, 7/30/99).

## 4.2 Diagnostics

An instrumentation section would be incorporated into the compressor section. It would comprise:

- Transverse emittance measuring wire scanners. A minimum of 3 need to be arranged with optimal betatron phase advance.
- Energy spread measurement using a single wire scanner located in the high dispersion region of the chicane.
- Orbit control with precision BPMS to correct the dispersion in the bunch compressor. These BPMS dictate that a small-diameter, moveable beam pipe be used in the chicane bends rather than a single large chamber to accommodate the variable chicane strength.

- Pulse-to-pulse measurement of beam position in the high-dispersion region to measure energy jitter.
- Precision phase measurement for pulse-to-pulse beam phase measurement and for monitoring long-term drifts in phase.
- A precision phase reference system to reference the above measurements.
- Synchrotron light monitor for streak camera measurements of bunch length.
- Microwave cavity resonators for monitoring bunch length.

The instrumentation section is as much necessary for tuning the bunch compressor as it is for accelerator physics experiments to study emittance growth phenomena associated with producing short bunches. Of equal importance is the exercise of operating the bunch compressor in the environment of the LCLS and solving the problems of phase stability in the linac RF system.

As has been detailed before, the majority of these components are available for reclamation from the SLC. There is a considerable cost benefit to this approach, as well as the advantage of using components that are already integrated into the SLAC accelerator control system.

# 5 Properties of the Undulator Radiation

The radiation wavelength, the average brightness and the peak brightness of the incoherent radiation in the LCLS reference design can be compared to the following cases:

Operation over a wider range of energies up to 50 GeV

Operation with the larger emittance and bunch length using the damping ring beam

Operation with a much shorter test section of the LCLS undulator

## 5.1 Wavelength

The wavelength of the incoherent radiation, $\lambda_r$, depends only on the undulator magnetic parameter, $K$, the undulator wavelength, $\lambda_u$, and the beam energy, as given by

$$\lambda_r = \lambda_u \left(1 + K^2\right) / 2\gamma^2$$

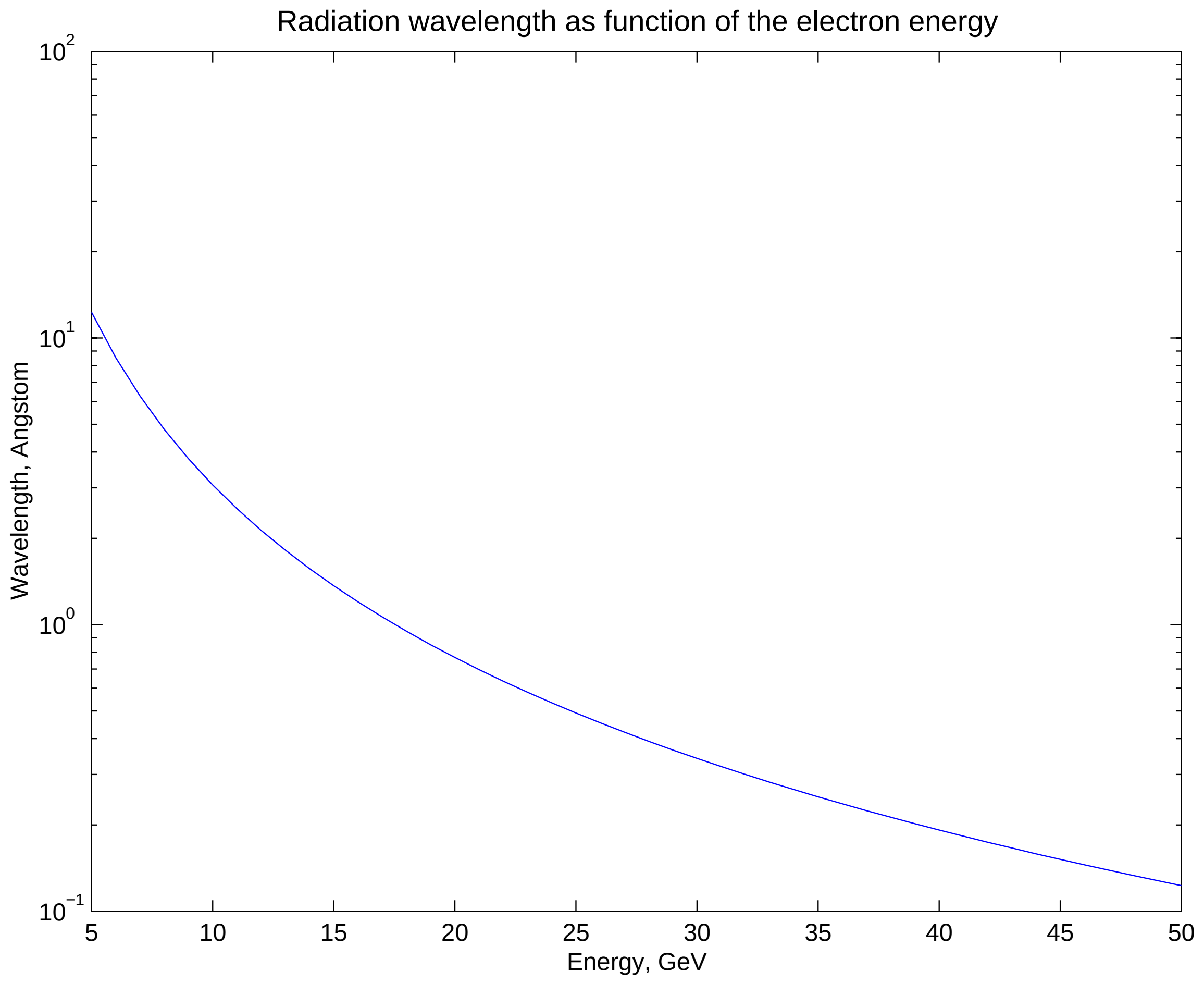

Figure 6: Wavelength of the spontaneous radiation as a function of the electron beam energy

The wavelength is independent of the emittance or bunch length. Figure 5 shows the quadratic decrease in wavelength attainable as the energy is raised above the 15 GeV LCLS design energy.

## 5.2 Brightness

The brightness of the radiation depends both on the undulator and beam properties. The noncoherent radiation brightness decreases quadratically with the larger emittance of the damping ring electron beam. There is also a decrease in brightness from using fewer undulator periods, for example, if shorter test sections of undulator were to be used. The brightness does increase with the higher energies achievable with the longer linac. The average brightness is a function of:

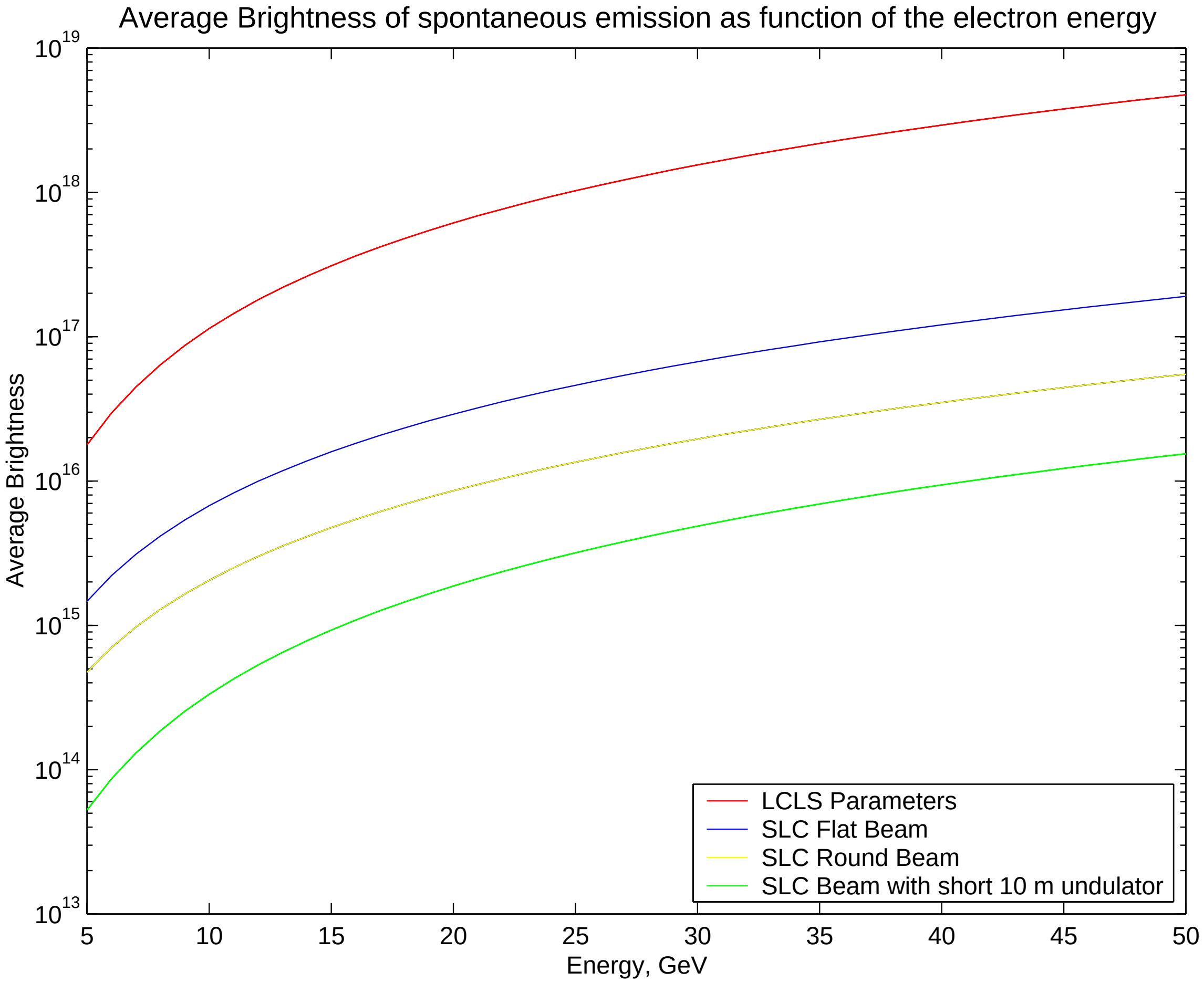

Figure 7: The average brightness of the spontaneous emission for the LCLS is compared to brightness obtained with an SLC flat (low vertical emittance) and round (equal emittances) beams and for an SLC

| | |
|---|---|
| $N_e$ | charge per bunch |
| $f_r$ | bunch repetition frequency |
| $\varepsilon_n$ | normalized transverse emittances |
| $\beta_{xy}$ | undulator beta functions |
| $\Delta\omega/\omega$ | radiation bandwidth |

and can be written

$$B_{av} \quad = \quad \frac{f_r N_e \alpha \dfrac{K^2}{1+K^2/2} F_1^2(K) 10^{-3}}{4\pi \quad \Sigma \qquad \Delta\omega/\omega}$$

where

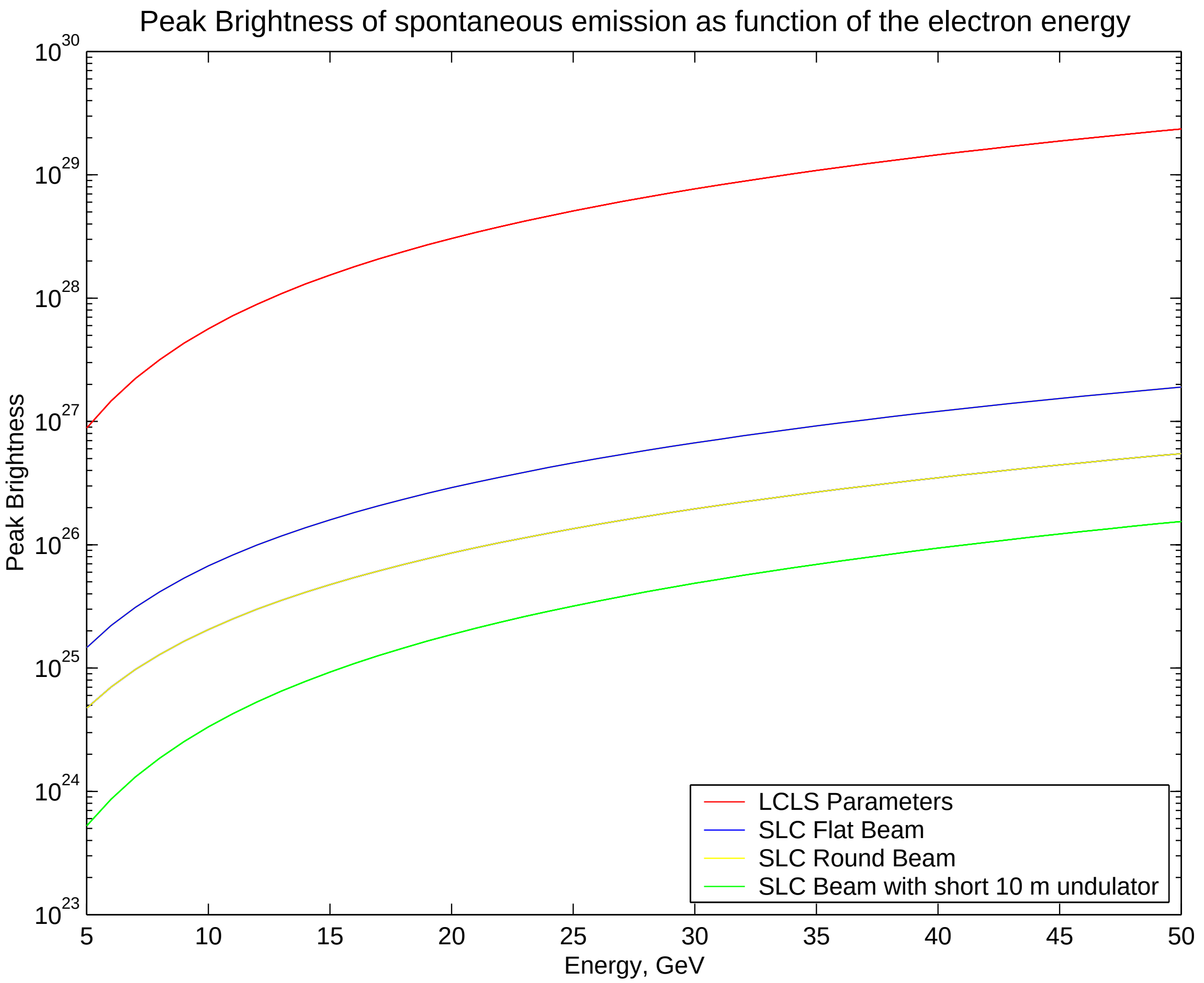

Figure 8: The peak brightness of the spontaneous emission for the LCLS is compared to brightness obtained with an SLC flat (low vertical emittance) and round (equal emittances) beams and for an SLC beam in a short test undulator.

$$\Sigma = \left\{\frac{\varepsilon_n \beta_{xy}}{\gamma} + \frac{\lambda_r \lambda_u N_u}{16\pi^2}\right\}\left\{\frac{\varepsilon_n}{\gamma\beta_{xy}} + \frac{\lambda_r}{\lambda_u N_u}\right\}$$

$$F_1(K) = J_1\left(\frac{K^2}{4(1+K^2/2)}\right) - J_0\left(\frac{K^2}{4(1+K^2/2)}\right)$$

$$\alpha^{-1} = \frac{2\varepsilon_0 hc}{e^2} = 137.05$$

In figure 7 the average brightness is shown as a function of energy for an electron beam with the LCLS design parameters compared to the parameters for the damping ring-linac beam. The damping rings can provide either round beams with equal emittances 0f 15 mm mrad, or flat beams with vertical and horizontal emittances of 2x30 mm mrad. The flat beams produce a brighter beam since the brightness shown in figure 4 depends on the product of the transverse emittances. The case for a shorter, test undulator is also shown.

The peak brightness of the radiation is also dependant on the electron beam bunch duration, τ, and is given by

$$B_{pk} = \frac{N_e \alpha \frac{K^2}{1+K^2/2} F_1^2(K) 10^{-3}}{4\pi\sqrt{2\pi}\,\tau \quad \Sigma \qquad \Delta\omega/\omega}$$

The peak brightness as a function of energy is shown in figure 8 for the LCLS case and the damping ring-linac beam.

The loss in brightness associated with the larger emittance of the damping ring can be partly compensated by employing a higher charge per bunch. Although the damping rings have been demonstrated to easily operate with the 5 nC per bunch listed in table 1, there will be a trade off in the linac emittance growth. More extensive simulation of the present linac would be required to find the optimum charge density above which linac wakefields would cause unacceptable emittance growth.

# 6 Future Options

The 30 GeV energy operation at 10 Hz is described here because it coincides with the current operating regime for the linac in its PEP II injection mode. Higher energies, up to 50 GeV are also available in the linac at the cost of powering more of the linac accelerating sections. The additional acceleration in sectors 20 through 30 has no effect on the PEP II injection. The FFTB beam lines are able to transport 50 GeV beams, giving the possibility of even shorter wavelength X-ray production in the LCLS. If the demand for higher average brightness becomes an issue for the experimentalist then the repetition rate of the facility can be raised from 10 Hz to 120 Hz, where the only major concern would be the additional cost for electrical operating power.

The capability to make shorter wavelength incoherent radiation may well remain a significant part of the LCLS experimental capability, in which case the sector-4 bunch compressor installation would be a permanent feature of the LCLS construction.

A further improvement in peak brightness could be achieved by exploiting the present low repetition rate of the damping rings and operating them at lower energy. With 30 Hz operation at 750 MeV versus the present 1.19 GeV we can expect a factor 4 improvement in transverse emittances as well as shorter bunch lengths (see, Proceedings of DR2000).

Looking further ahead it is conceivable that a second rf photo-injector would eventually be built near the beginning of the linac to provide high-brightness, high-energy beams to produce coherent radiation at the shorter wavelengths. The sector-4 bunch compressor would be an integral part of this upgrade.

# 7 Conclusion

The addition of a second bunch compressor to the SLAC linac at the 4.5 GeV location gives a capability of delivering bunches as short as 100 μm with peak currents possibly as high as several kilo Amperes. The mode of operation where 30 GeV beams are delivered LCLS/FFTB is fully compatible with the present PEP II injection and entails no additional operating costs for the linac. The FFTB beam line handles beams up to 50 GeV so there is also an option for even shorter wavelength X-ray undulator radiation using the full complement of linac klystrons.

Such a facility provides an intermediate source of Xray radiation with a brightness between that of present facilities and the proposed LCLS project. This would provide the laboratory with an opportunity to develop techniques in handling and diagnosing high-power X-ray beams. Tests for future LCLS experiments, such as laser pump probe measurements, could be performed with the high-brightness spontaneous radiation from this facility.

The experience gained in the production and tuning of short bunch electron beams is also an invaluable intermediate step towards the LCLS. The compressor components would be exact prototypes for the LCLS and would be operating with the same linac and in the same RF environment. The machine physics associated with operating the linac with short bunches is sufficiently new and challenging to be of interest to the NLC and other future accelerator studies at SLAC.

# 8 References[1]

[1] A copy of this note and the following references may be found at the following URL http://www.slac.stanford.edu/~pkr/home.html